\documentclass[letterpaper]{article}
\usepackage{pervasives}

\title{Making High-Performance Robots Safe and Easy to Use\\
For an Introduction to Computing}
\author{Joseph Spitzer\textsuperscript{\rm 1},
  Joydeep Biswas\textsuperscript{\rm 2}, and
  Arjun Guha\textsuperscript{\rm 1} \\
University of Massachusetts Amherst\textsuperscript{\rm 1} and
University of Texas at Austin\textsuperscript{\rm 2}
}

\begin{document}

\maketitle

\begin{abstract}
Robots are a popular platform for introducing computing and artificial
intelligence to novice programmers. However, programming state-of-the-art
robots is very challenging, and requires knowledge of concurrency, operation
safety, and software engineering skills, which can take years to teach. In this
paper, we present an approach to introducing computing that allows students to
safely and easily program high-performance robots. We develop a platform for
students to program \emph{RoboCup Small Size League} robots using JavaScript.
The platform 1)~ensures physical safety at several levels of abstraction,
2)~allows students to program robots using JavaScript in the browser, without
the need to install software, and 3)~presents a simplified JavaScript semantics
that shields students from confusing language features. We discuss our
experience running a week-long workshop using this platform, and analyze over
3,000 student-written program revisions to provide empirical evidence that our
approach does help students.
\end{abstract}

\section{Introduction}

Robots are frequently used to introduce computing to novice
programmers~\cite{Osborne:2010,Featherston:2014,Magnenat:2014,Dee:2017,Gucwa:2017,Musicant:2017,Paramasivam:2017,Doran:2018,Paspallis:2018}.
Since they are a hands-on medium, it is relatively easy to make an engaging
robotics-based curriculum. Moreover, robots are a natural platform to introduce
a variety of core STEM subjects, including geometry, mechanics, and
programming. However, many robotics-based curricula use robots that are
designed for education, and do not represent the state-of-the-art in
robotics~\cite{Osborne:2010,Featherston:2014,Magnenat:2014,Dee:2017,Gucwa:2017,Musicant:2017,Estrada:2017,Doran:2018,Paspallis:2018}.

Our research group maintains a team of soccer-playing robots that compete in
the \emph{RoboCup Small Size League (SSL)} tournament~\cite{RC}. We designed,
built, and programmed these robots ourselves, and wanted to use them in an
outreach workshop for several reasons. 1)~SSL robots are compelling because
they exhibit high-performance behaviors and demonstrate what state-of-the-art
robots can do. For example, they are omnidirectional, move at high speeds, and
can manipulate a ball rapidly. 2)~We can explain and demonstrate how the robots
were built and the equipment needed to do so, namely CNC mills and CAD tools.
3)~SSL robots are designed to compete in teams, thus students can use them to
write cooperative and competitive agents for soccer and other activities (\eg
robot tag or navigation in a maze). 4)~Soccer is a well-known domain.

However, high-performance robots such as ours, are not well-suited for novice
programmers. 1)~If programmed incorrectly, the robots can cause injury because
they are able to move very rapidly. 2)~We program the robots using a
sophisticated low-level C++ API that lends itself to writing high-performance
code, but requires significant expertise to use. 3)~For competitive play, the
robots employ a multi-tier abstraction for high-level team-wide planning and
low-level control that continuously updates the robots'
roles~\cite{Browning:2004}, which makes it difficult to issue simple movement
commands to an individual robot.

\begin{figure}
  \centering
  \includegraphics[width=.95\columnwidth]{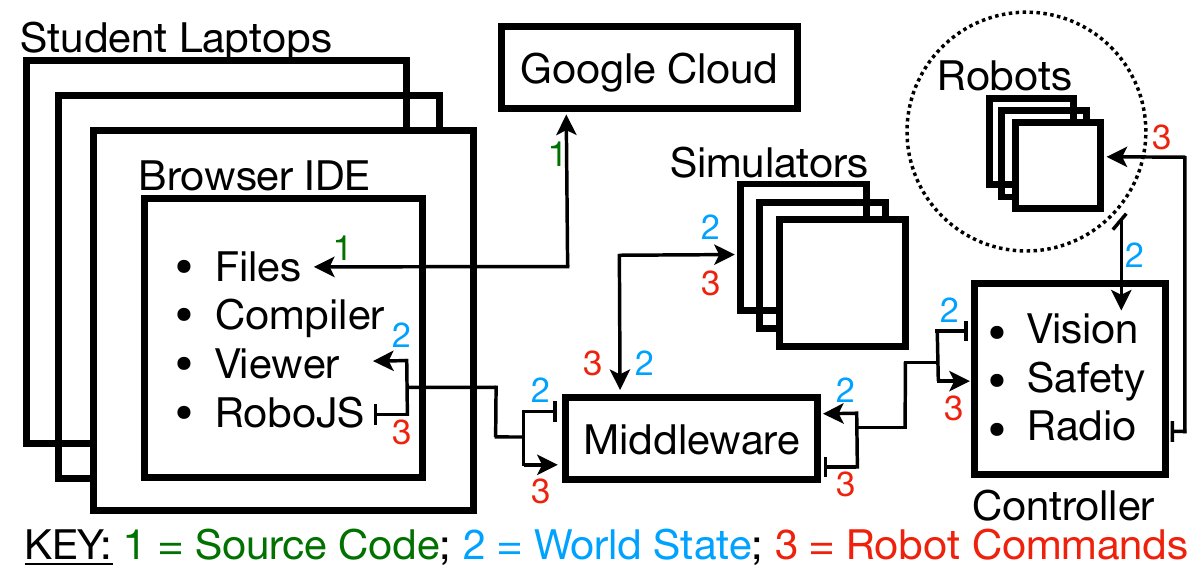}
  \caption{A schematic representation of the major components
  and data flow in our system.}
  \label{schematic}
\end{figure}

\paragraph{Contributions} The diagram in \cref{schematic} depicts the design of
our system, which addresses these challenges.

\begin{enumerate}
  \item We develop a robot control platform that checks for safety at several
  layers of abstraction, which allows novices to safely work with
  high-performance RoboCup SSL robots.

  \item We develop a simplified programming stack for controlling robots over
  the network (\emph{Middleware} in \cref{schematic}). This programming stack
  is resilient to transient network faults and provides simple interfaces to
  low-level robot motion controllers that provide functionality like grid
  navigation and ball interception.

  \item Although we use a professional programming language, and not a
  programming language designed for education, we take several measures to
  shield students from complex and confusing behavior that the language
  normally exhibits. Specifically, we use JavaScript and build a
  source-to-source compiler that simplifies the semantics of JavaScript to
  make it more intuitive.

  \item We develop a curriculum for a week-long outreach workshop that
  gradually introduces students to progressively more advanced mathematics
  and more sophisticated forms of robot control. Our robot programming APIs
  have multiple layers of abstraction that support this curriculum design.
\end{enumerate}

We discuss our experience running a week-long workshop using this platform,
and analyze over 3,000 program revisions created over the course of the
workshop to argue that our approach does help students.

The rest of this paper is organized as follows. \Cref{related-work} reviews
related work. \Cref{robotics-platform} presents our robot programming platform
and our approach to safety. \Cref{novice-js} presents our JavaScript
programming environment, and the approach we use to simplify JavaScript for
novice programmers. \Cref{workshop} presents our workshop curriculum.
\Cref{eval} presents evidence that our programming environment helps students
catch bugs. Finally, \cref{conclusion} discusses future work and concludes.

Our software can be found on GitHub within the \texttt{ut-amrl/robo-js}
repository.

\section{Related Work}
\label{related-work}

Young children exposed to basic computer science concepts with robots
exhibit promising levels of comprehension by way of such a hands-on
medium~\cite{Magnenat:2014,Martinez:2015}. These sessions---and others
including \cite{Featherston:2014}---use a block-based programming
language alongside a simplified robotics platform.
Scratch\footnote{\url{https://scratch.mit.edu}},
Alice\footnote{\url{https://www.alice.org}}, and
Blockly\footnote{\url{https://developers.google.com/blockly}} are some of
the most common block-based programming languages. Lego
Mindstorms\footnote{\url{https://www.lego.com/mindstorms}},
Thymio\footnote{\url{https://www.thymio.org}}, and
Linkbot\footnote{\url{https://www.barobo.com}} are popular educational robots;
for each there is support of both block and text-based programming,
either natively or through a third-party.

In contrast to other outreach workshops and exploratory activities within the
high-school age range~\cite{Osborne:2010,Musicant:2017,Paspallis:2018}, our
students do not build or augment robotics hardware. All our robots were created
independently; on-site setup and maintenance was performed by the workshop
instructors.

Our formalization of a layer-based abstraction appears to be a novel design for
short-term computer science outreach. There is an analogous approach outlined
on a much different scale; a university curriculum~\cite{Doran:2018}. Students
begin with Java in CS1 using Lego Mindstorms by way of
leJOS\footnote{\url{http://www.lejos.org}}, a Java based replacement for the
existing Lego firmware. In future courses, native Lego hardware is replaced with
an Arduino and Raspberry Pi for use of common libraries in C, Java, and Python.

Using simulation in conjunction with physical robots is quite common.
Industrial grade simulation, provided by
Gazebo\footnote{\url{http://gazebosim.org}}, is excessive for introductory use
as it can easily overwhelm students. For this reason, simulators wherein the
target audience is users new to CS have been developed into products such as
Robot Virtual Worlds\footnote{\url{http://www.robotvirtualworlds.com}} and
RoboSim\footnote{\url{https://c-stem.ucdavis.edu/studio}}. The latter can be
used seamlessly with physical robots~\cite{Gucwa:2017}. They outline ideal
principles of a simulated environment, including uniform code and accurate
real-world representation.

We use a modification of the standard vision system~\cite{SSLVision} of RoboCup
SSL. There exists an implementation of a similar lower-end system to interface
with Lego Mindstorms using a web-cam and sending commands over
Bluetooth~\cite{Estrada:2017}.

To our knowledge, the most similar work to date is that done by a group from
the University of Washington~\cite{Paramasivam:2017}. They conducted a
week-long outreach workshop for high school students using JavaScript.
Providing a small API for robot control, students wrote programs in a
browser-based editor for a
TurtleBot\footnote{\url{https://www.turtlebot.com}}. Perhaps the biggest
difference was the robots that were used. The standalone TurtleBots with a
touchscreen and speech input-output supplied a platform for programs centered
around human-robot interaction. By contrast, our RoboCup SSL robots are in use
simultaneously from the afternoon of the first day, thus activities are rather
focused on multi-agent systems.

RoboCup serves as a challenging test-bed for several research problems,
including failure recovery~\cite{srtr}, time-optimal control~\cite{tsocs}, and
multi-agent planning~\cite{cap}. In particular, RoboCup Jr.~\cite{robocupjr}
serves as a bridging league where high-school students can compete at
international RoboCup competitions. Our work continues the tradition of using
RoboCup to inspire the future generation of robotics by catering specifically
to students via introduction to computing workshops.

\section{RoboCup Robots for Novice Roboticists}
\label{robotics-platform}

We present our state-of-the-art soccer-playing robots and the associated
software platform, including the existing design and the necessary changes we
made for the workshop. These include increased safety measures, middleware to
tie the low-level robotics hardware layer with the end-user environment, and
student access to our auxiliary simulator.

\subsection{Hardware Platform}

\begin{figure}
  \centering
  \includegraphics[width=.95\columnwidth]{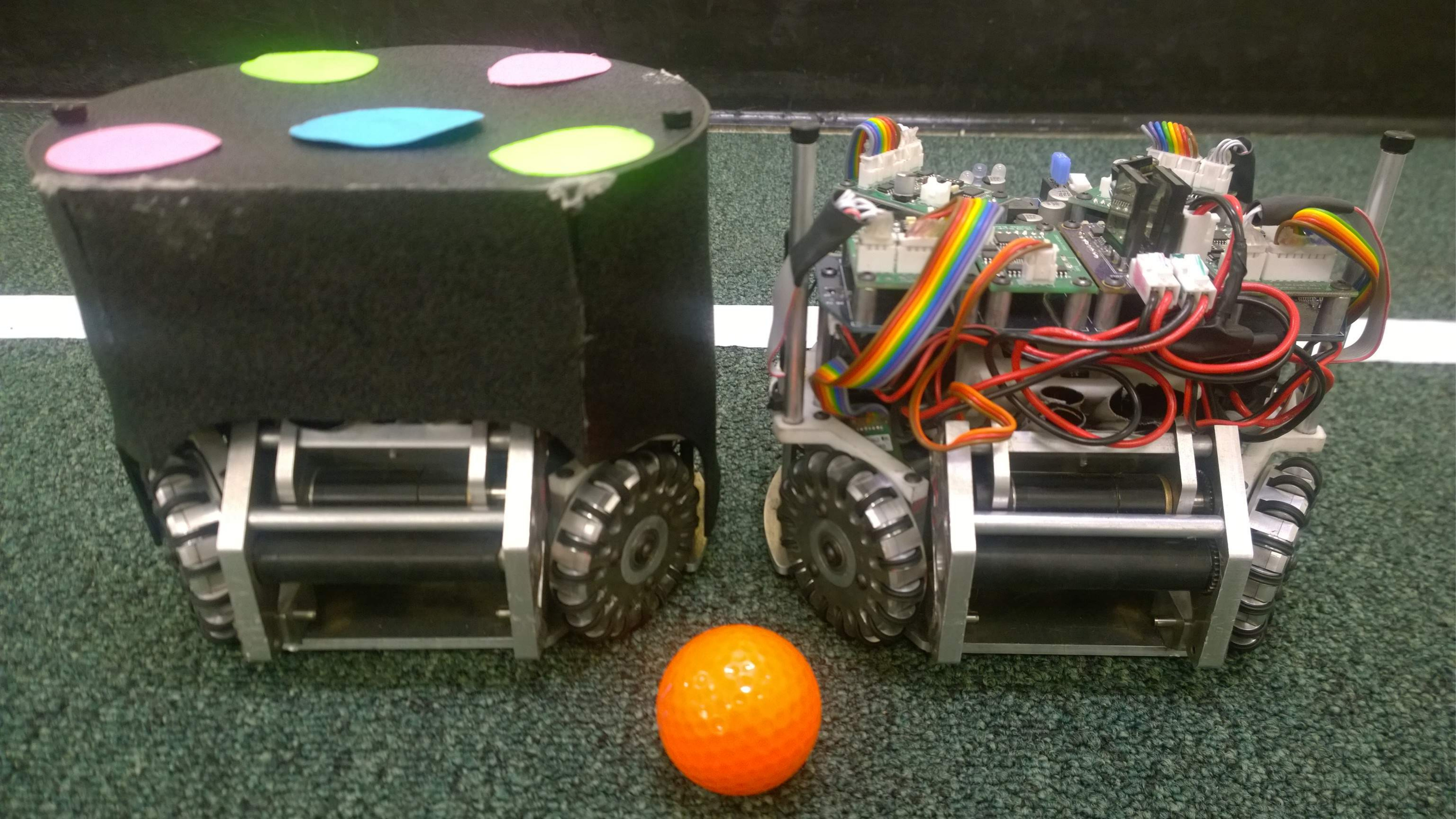}
  \caption{One of the robots alongside another with its casing removed.
    A colored golf ball is used as the ball.}
  \label{the-robots}
\end{figure}

\Cref{the-robots} shows our custom-designed robots that we built to compete in
the RoboCup SSL. The robots fit within a cylinder of 18cm in diameter and are
15cm tall. The electronics system is modular with the main circuit board
handling receiving and processing commands as well as power distribution.
Auxiliary boards control wheel motors and kick behavior. Kicking is
accomplished by accelerating a ferromagnetic kicking rod using a custom-wound
electromagnet with energy stored in a bank of capacitors. One of the immediate
fascinations of these robots is that their movement is omnidirectional, and
they are capable of moving precisely at high speeds while manipulating the ball
in motion.

All robots move within a calibrated field, which is simply a Cartesian plane,
thereby permitting use of standard Euclidean geometry and algebra in
calculations. The field dimensions at our venue were 2.4m by 3.6m. An
overhead camera can uniquely identify each robot by their top dot
pattern~\cite{SSLVision}. This vision system can instantaneously observe robots
and a ball on the field; positions and velocities are updated in real-time.

\subsection{Robot Software Stack}

The software platform is written in C++, with networking done over UDP
multicast. It is multi-threaded in order to support concurrent
\begin{inparaenum}[1)]
  \item state estimation to update the state of the world based on vision data,
  \item control and planning to assign roles to agents and to compute safe
  collision-free control trajectories, as well as
  \item collation and transmission of commands to the robots.
\end{inparaenum}

Multi-robot coordination and task allocation is performed using
Skills-Tactics-Plays (STP)~\cite{Browning:2004}. \emph{Skills} define
low-level actions, which the physical robot can perform, such as movement or
kicking. \emph{Tactics} group skills and determine which one to execute at a
given time. For example, a \emph{Goalie} tactic will try to block the ball if
the trajectory of the ball would land in its goal. Finally, \emph{Plays} group
tactics for synchronized activity among multiple robots. Examples of plays
include \emph{Kickoff} or \emph{Free-kick}.

We add a single tactic named \emph{Marionette} to the STP stack running on our
soccer robots. This tactic is special in the sense that it simply acts on data
received from outside the system, after applying safety checks. It has the
skills to move, dribble, catch, block, and kick. These skills are sufficient
for the workshop and have corresponding library methods for the end programmer
to interface with. For some skills there are multiple library methods exposed
to the students; see \cref{diff-move-cmds}. Given the nature of the Marionette,
there is no need for plays as all robots only run this tactic.

Also, we republish the current world state to UDP so it may be consumed outside
the platform. These two items; the addition of the Marionette tactic and the
modification to expose system state were the two main changes we had to make.
In this sense we leverage all existing capabilities of the system by providing
a means to exchange data, through our application middleware. This middleware
serves as the intermediary between the low-level C++ code that runs on the
robot controller and the high-level JavaScript code the students write in the
browser.

\subsection{Robot Safety Measures}

Given these are high-performance robots it is important that safety be taken
into account both for the students and the robots. As such, we employ the
following safety measures:
\begin{enumerate}

\item \textbf{Reduced Motion Model:} To ensure safe gameplay on a reduced field
size, we limited the maximum speed of the robots to 1m/s (the robots are
capable of maximum speeds in excess of 4m/s).

\item \textbf{Crash Prevention Buffer:} Each robot has a uniform safety margin
that surrounds it, enforced by Dynamic Safety Search~\cite{dss}. This margin
accounts for multi-robot dynamics and prevents robots from running into each
other, even if the student-written programs erroneously command them to do so.

\item \textbf{Command Timeouts:} The robots have a command timeout such that if
the robot control stack does not receive any commands from the student programs
within a sliding five second window, the control stack aborts any active
actions and brings the robots to a halt. Execution resumes seamlessly with the
next command received.

\item \textbf{Robot Identification Number} Before any robot commands may be
issued, the programmer must explicitly set the number of a robot that is
presently available for control, thus disallowing the program from commanding
robots that are not on the field.

\item \textbf{Field Boundaries} If a statement in a student's program
erroneously instructs the robot to go beyond the field boundaries then the
Marionette tactic truncates the commanded locations to allowable value(s).

\item \textbf{Skill-based Checks} The Marionette tactic verifies that the
low-level skill commanded by the student program is indeed applicable based on
the current world state -- this allows it to catch errors where the student
program for example tries to kick the ball when it is far from the robot or not
in front of the robot.
\end{enumerate}

\subsection{A Simulation Viewer in the Browser}
\label{sim}

\begin{figure}
  \centering
  \includegraphics[width=.95\columnwidth]{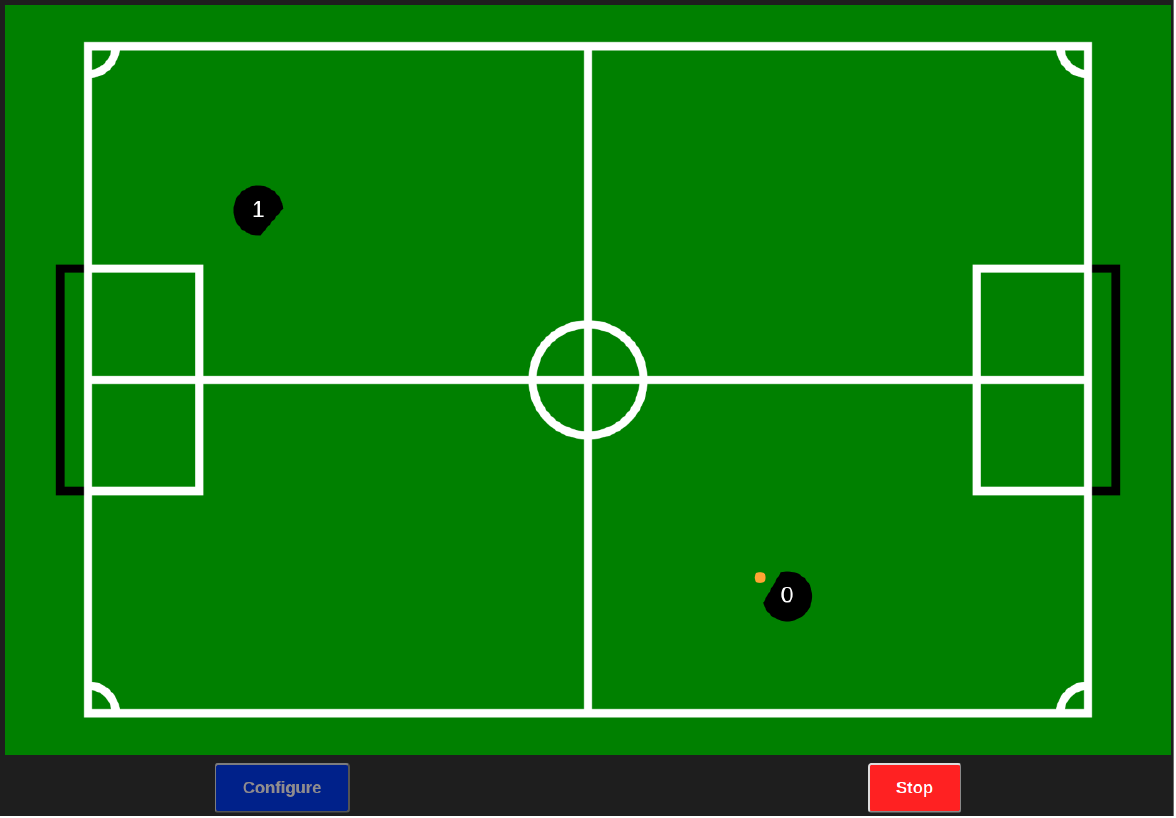}
  \caption{Two robots passing in simulation.}
  \label{sim-pic}
\end{figure}

We provide a web-based viewer for corresponding simulators hosted on remote
servers. The viewer displays the world state, relayed to it by way of the
middleware intermediary in real-time. Beneath the visual pane are basic
configuration options. Based on the programming activity, the students can
select initial simulator conditions from a supplied list. Such items vary the
number of robots, the ball and robot positions, and in some cases the
background (\eg soccer pitch or a maze). Some configurations involve
randomization, so to provide a better means for program testing. \Cref{sim-pic}
shows our viewer while a simulation is running; the animations are 2D from a
top-down perspective and are made using the HTML Canvas.

The choice to provide simulator viewing and control in the browser is twofold.
For one it allows us to easily abstract away the exhaustive detail of the
native viewer and the rigid command-line interface of the simulator. Also, by
using the browser, the need for installation and hardware requirements is
alleviated.

\section{JavaScript for Novice Programmers}
\label{novice-js}

In this section, we first discuss why JavaScript seems to be a reasonable
language to introduce computing to novices. We then show that JavaScript has
several counterintuitive features that make it a poor choice for teaching
beginning programmers. We then present \emph{RoboJS}, a source-to-source
compiler for JavaScript that transparently simplifies its semantics, which
makes it behave like a ``normal'' language. Finally, we outline our end-user
robotics library and how these pieces come together in the RoboJS IDE.

\subsection{Why JavaScript?}

There are several reasons why JavaScript is a good choice for an introduction
to computing. First, JavaScript is one of the most widely used programming
languages in the world~\cite{GH-lang-classification,TIOBE,SO-survey}, thus it
introduces students to a technology that they are likely to encounter again.
Second, there are several web-based IDEs for JavaScript that allow students to
write and run programs within a web browser, without the need to install any
further software. Finally, since JavaScript is a dynamically-typed language,
there is no need to teach students how to work with a type-checker, which is
necessary to program in Java or C.

\subsection{The Case Against JavaScript}

JavaScript has several peculiar features, which can confound novice and expert
programmers~\cite{maffeis:jssemantics,guha:js}. In this section, we present a
few of these features and argue that they make JavaScript a poor choice for
teaching programming.

\paragraph{No Arity Mismatch Errors} JavaSript does not have arity-mismatch
errors, thus it is not an error to call a function with too many or too few
arguments. If a function receives too many arguments, the extra arguments are
silently dropped. If a function receives too few arguments, the elided
arguments are set to the special value \lstinline|undefined|. For example,
our robot programming API has a \lstinline|moveTo| function that takes three
arguments: an $x$-coordinate, a $y$-coordinate, and an angle. We found that
students frequently forget to supply the angle argument, and would call
\lstinline|moveTo| with only the coordinates. In most other languages, this
omission would trigger an error (either during compilation or at runtime).
However, no error occurs in JavaScript.

\paragraph{Implicit Type Conversions} JavaScript performs many implicit type
conversions under the hood. Whereas some type conversions can be valuable,
others are very counterintuitive. For example, the relational operators of
JavaScript convert \emph{all} values to numbers, including functions and
objects. Functions and objects are converted to \lstinline|NaN|, which is a
floating-point number, and all comparisons with \lstinline|NaN| produce
\lstinline|false|. In our robot programming API there is a function called
\lstinline|getBallPosX| that takes zero arguments and returns the current
position of the ball. Students would frequently forget to call zero-argument
functions using parenthesis, and write code such as the following:
\begin{lstlisting}
if (robot.getBallPosX > 0) { ... }
\end{lstlisting}
The code above is wrong, since it is comparing the function, and not the result
of calling the function. However, it does not raise an error. In other
languages, including dynamically-typed languages such as Python, the equivalent
code would raise an error. A mistake such as this is very hard to identify,
even for experts.

\paragraph{Non-Existent Fields} In JavaScript, it is not an error to access
a non-existent field from an object. When a program does so, JavaScript returns
the special value \lstinline|undefined|. Worse, a program can explicitly set
a field to the value \lstinline|undefined|, which makes it even harder to
discern if the field exists or not. Due to this behavior, simple spelling
mistakes can produce \lstinline|undefined|. In contrast, accessing
non-existent fields produces an error in Java (during compilation) and Python
(at runtime).

\subsection{A Comprehensible Subset of JavaScript}

We develop a subset of JavaScript that we call RoboJS, which eliminates the
problems listed above (among others). Moreover, our IDE transparently ensures
that students' programs are in RoboJS, using a source-to-source JavaScript
compiler that runs in the browser. The RoboJS compiler uses static checks when
possible to catch errors before the program runs. When static checks are
infeasible, it inserts dynamic checks that execute at runtime. Therefore, when
a user uses a forbidden language feature (knowingly or not), it produces an
error and aborts the program. The compiler is careful to ensure that its
operation is transparent: the user never sees the inserted checks and the
source locations are preserved so that the inserted checks do not affect how
other errors are reported. Finally, since RoboJS is a strict subset of
JavaScript, any RoboJS program runs outside the IDE.

For instance, we have seen that the comparison operators in JavaScript, such as
the greater-than operator, perform unintuitive implicit type conversions. In
contrast, RoboJS requires both arguments of the greater-than operator to be
numbers. To do so, it replaces all occurrences of \lstinline|x > y| in
students' code with a call to a library function that ensures that $x$ and $y$
are numbers (\lstinline|checkedGT(x, y)| in \cref{checkedGT}). Note that
students never see this function nor know that their code is being rewritten
in this manner.

\begin{figure}
  \centering
  \scriptsize
  \begin{lstlisting}
function checkedGT(lhs, rhs) {
  if (typeof lhs !== 'number' || typeof rhs !== 'number') {
    throw Error('Arguments of ">" must both be numbers.');}
  return lhs > rhs;}
  \end{lstlisting}
  \caption{A simplified version of the runtime check for the greater-than
  operator in RoboJS.}
  \label{checkedGT}
\end{figure}

Similarly, in RoboJS, a function must receive exactly as many arguments as it
declares. To accomplish this, the RoboJS compiler adds a check to the start of
every function to assert that the number of actual arguments
(\lstinline|arguments.length|) is equal to the number of declared arguments. If
not, RoboJS produces an error. Again, this check is transparent to the
programmer.

RoboJS has several other checks and carefully constructed error messages, which
shield students from JavaScript's peculiar features. In \cref{eval}, we present
evidence to show that RoboJS is effective and does help students catch mistakes
early.

\subsection{Programming Robots with JavaScript}
\label{diff-move-cmds}

In addition to shielding students from JavaScript's peculiar semantics, RoboJS
also allows students to control robots using JavaScript. The RoboJS API is
designed to gradually increase the amount of control that students have over
the robots. For example, RoboJS provides several different functions to move
robots (\cref{large-code-example}). At the beginning of the workshop, we
introduced students to basic motion commands that moved the robots on a
discrete grid (\cref{large-code-example}, \cref{easy}). This allowed students
to get comfortable typing JavaScript code, without the need to understand
geometry in detail. Subsequently, we had students use functions that moved the
robot by a given number of cells (\cref{large-code-example}, \cref{medium}).
This gave the students experience with function parameters and helped reduce
repetitive coding. Finally, we presented movement in a continuous environment
with angles of rotation, absolute motion, and relative motion
(\cref{large-code-example}, \cref{hard}). By gradually introducing students to
more advance features as the workshop progressed, we made an effort not to
overwhelm them with too much new content at once.

\begin{figure}
  \centering
  \scriptsize
  \begin{lstlisting}
// Beginner move commands:#\label{easy}#
robot.moveForward(); robot.turnLeft(); robot.turnRight();

// Intermediate move commands:#\label{medium}#
robot.moveByXCells($cellsX$); robot.moveByYCells($cellsY$);

// Advanced move commands:#\label{hard}#
robot.moveByX($distanceX$); robot.moveByY($distanceY$);
robot.moveByXY($distanceX$, $distanceY$); robot.turnBy($degrees$);
robot.moveBy($distanceX$, $distanceY$, $degrees$);

robot.moveToX($x$); robot.moveToY($y$); robot.moveToXY($x$, $y$);
robot.turnTo($degrees$); robot.moveTo($x$, $y$, $degrees$);
  \end{lstlisting}
  \caption{Progression of precision and detail with respect to how the students
  programmed robot movement.}
  \label{large-code-example}
\end{figure}

\begin{figure}
\begin{subfigure}{\columnwidth}
  \centering
  \scriptsize
  \begin{lstlisting}
robot.moveToXY(100, 100, function() {
  robot.turnTo(180, function() {
    console.log('Done');
  });});
  \end{lstlisting}

  \caption{Programming a robot using callback functions.}
  \label{callback}
\end{subfigure}

\begin{subfigure}{\columnwidth}
  \centering
  \scriptsize
  \begin{lstlisting}
robot.moveToXY(100, 100);
robot.turnTo(180);
console.log('Done');
  \end{lstlisting}
  \caption{Programming a robot with RoboJS.}

  \label{straight-line}
\end{subfigure}

\caption{JavaScript does not support blocking I/O, which can make it hard to
write programs that communicate with any external service, including our robots.
However, RoboJS simulates blocking I/O, which makes programming easier.}
\end{figure}

Another aspect of the library methods are the lack of data structures.
Everything is a primitive value, both as input to the methods, and output
from them. We only taught the students about numbers, booleans, and strings.
For instance, instead of a single method to return the position of a robot in
the continuous plane there were three; one each for the $x$-coordinate,
$y$-coordinate, and angle. This API design allows us to avoid introducing
arrays and objects.

\subsection{An In-Browser IDE and Runtime for RoboJS}

RoboJS includes an IDE that runs in a browser. The RoboJS IDE has a
straightforward interface: 1)~a handful of buttons, including buttons to
start the program, stop the program, and open the robot simulator (\cref{sim}),
2)~a list of files, 3)~a code editor based on
Monaco\footnote{\url{https://microsoft.github.io/monaco-editor}}, and 4)~a
REPL where students can view textual output and execute simple expressions
immediately.

The RoboJS runtime system shields students from a key limitation of web browser
APIs, which is that all network I/O is non-blocking. Thus code that makes a
network request must use a callback function to receive the response. Since our
platform relies on the network to issue commands to robots, it appears that we
need to teach callbacks and higher-order functions to have students do anything
of interest. For example, if we wanted to have a robot first move to a location
and turn only after the move command completes, we would have to teach students
to write nested functions (\cref{callback}). This style of programming is not
suitable for beginners.

Instead, the RoboJS runtime allows students to write simple, straight-line
programs (\cref{straight-line}). Its APIs simulate blocking I/O on top of the
web browser's non-blocking primitives by building on a tool called
Stopify~\cite{Baxter:2018}. Stopify is a source-to-source JavaScript compiler
that simulates multiple, cooperative threads of execution in
JavaScript.\footnote{Technically, Stopify simulates first-class continuations,
which we use to simulate cooperative threads~\cite{wand:multiprocessing}.}
Using Stopify, RoboJS maintains two logical threads: one thread runs the user's
program and the other performs I/O operations. To perform an I/O operation,
the user thread sends a message to the I/O thread, and then suspends itself.
When the I/O operation completes, the I/O thread wakes the sleeping user thread
and sends it the result.

\begin{table}
  \centering
  \scriptsize
  \begin{tabular}{|r|p{7cm}|}\hline
    \thead{Day} & \thead{Summary} \\\hline
    \textbf{1} & \textbf{Topics:} Introduction to the robots and programming
    environment by example. Statements, variables, conditionals, and loops.
    Actuation, inputs, and sensing. \textbf{Activities:} Mazes with multiple
    open repetitive paths with and without dynamic obstacles.\\\hline
    \textbf{2} & \textbf{Topics:} Discrete coordinate geometry and relative
    motion. Functions. Planning and robot hardware. \textbf{Activities:} Fixed
    item collection based on a random robot location with and without
    adversarial obstacles.\\\hline
    \textbf{3} & \textbf{Topics:} Continuous coordinate geometry and absolute
    motion. Multi-agent robotics, navigation, and more on planning (\ie
    dynamic and adversarial). \textbf{Activities:} Robot tag, both chasing and
    avoiding agents to compete amongst each other and against supplied
    programs.\\\hline
    \textbf{4} & \textbf{Topics:} More geometry, with special attention to
    angles. Feature and API exposure needed to interact with the ball and play
    soccer. \textbf{Activities:} Penalty shootout, both goalie and striker
    agents to compete amongst each other and against supplied programs.\\\hline
    \textbf{5} & \textbf{Topics:} \emph{No new material.}
    \textbf{Activities:} 2v2 soccer with 4 distinct roles; goalie,
    defender, primary and secondary attacker. Students worked in
    teams to write these agents, with elementary default code provided.\\\hline
  \end{tabular}
  \caption{An overview of the topics and activities for each day of the
  workshop.}
  \label{day-summary}
\end{table}

\section{Computing and Robotics Workshop}
\label{workshop}

We ran a week-long workshop with twelve high-school students. The only prior
programming experience in this group was that one student was familiar with
TI-Basic and several had worked with Lego Mindstorms using a block-based
language. \Cref{day-summary} summarizes the curriculum and programming
activities that we used over the course of the week. The curriculum
incrementally introduces more sophisticated concepts in geometry, which was
supported by the layers of abstraction in the RoboJS API. The students started
every activity by programming in simulation. However, we encouraged them to
frequently test their programs on physical robots (with modest supervision by
the course staff). In the early activities that use discrete coordinates,
programs reliably behave the same in simulation and on the real robots. However,
with continuous coordinates and more sophisticated behaviors, such as catching
and passing a ball, there are inevitable discrepancies between the simulation
environment and the physical environment. This is an important lesson to learn
when programming robots; it is necessary to frequently test programs on
\emph{actual} robots.

\section{Evaluation of RoboJS}
\label{eval}

\begin{table}
  \centering
  \scriptsize
  \begin{tabular}{|r|r|r|r|r|r|}\hline
    \thead{Account} & \thead{Lines (L)} & \thead{Revisions (R)} &
    \thead{Files (F)} & \thead{L/R} & \thead{R/F} \\\hline
    1 & 7,591 & 180 & 23 & 42.2 & 7.8\\\hline
    2 & 6,478 & 208 & 14 & 31.1 & 14.9\\\hline
    3 & 3,795 & 164 & 26 & 23.1 & 6.3\\\hline
    4 & 7,871 & 296 & 25 & 26.6 & 11.8\\\hline
    5 & 4,227 & 196 & 24 & 21.6 & 8.2\\\hline
    6 & 5,253 & 194 & 18 & 27.1 & 10.8\\\hline
    7 & 13,598 & 508 & 32 & 26.8 & 15.9\\\hline
    8 & 18,796 & 462 & 29 & 40.7 & 15.9\\\hline
    9 & 23,087 & 590 & 29 & 39.1 & 20.3\\\hline
    10 & 15,472 & 432 & 17 & 35.8 & 25.4\\\hline
    \thead{Total} & 106,168 & 3,230 & 237 & 32.9 & 13.6\\\hline
  \end{tabular}
  \caption{An aggregate summary of program data per student account, which
  also shows the approximate program size (\ie lines per revision) and number
  of edits (\ie revisions per file).}
  \label{data-summary}
\end{table}

In this section, we evaluate the effectiveness of RoboJS by studying the kinds
of errors that RoboJS caught during our workshop. We leverage data from our
web-based IDE which saves a history of edits for every file, so that students
can easily revert to an older version if needed. Specifically, if a file has
changed since the last run, then the IDE saves a copy of the program to the
cloud when a user tries to run it again. Therefore, we expect that most
revisions are programs that should have (partially) worked. We gathered 3,230
revisions total across ten student groups (\ie accounts) over the course of
one week. \Cref{data-summary} summarizes the number of lines, revisions,
files, and their sizes across all accounts.

\begin{table}
  \centering
  \scriptsize
  \begin{tabular}{|r|r|r|r|}\hline
    \thead{Account} & \thead{Syntax Errors} & \thead{RoboJS Errors} &
    \thead{Revisions}\\\hline
    1 & 10 & 82 & 180\\\hline
    2 & 26 & 28 & 208\\\hline
    3 & 6 & 12 & 164\\\hline
    4 & 39 & 43 & 296\\\hline
    5 & 18 & 46 & 196\\\hline
    6 & 29 & 10 & 194\\\hline
    7 & 22 & 60 & 508\\\hline
    8 & 49 & 89 & 462\\\hline
    9 & 80 & 54 & 590\\\hline
    10 & 45 & 205 & 432\\\hline
    \thead{Total} & 324 & 629 & 3,230\\\hline
  \end{tabular}
  \caption{A count of JavaScript syntax errors and RoboJS runtime errors per
  student account. Note the RoboJS errors are estimated; see \cref{rt} for
  context.}
  \label{error-summary}
\end{table}

\subsection{JavaScript Syntax Errors}

JavaScript reports syntax errors itself and our IDE gives a modicum of
feedback: it highlights matching brackets and parenthesis. Therefore, we first
checked the syntax of all revisions and found that 10\% of the revisions had
syntax errors (\Cref{error-summary}). We found far fewer errors on the last day
of the workshop, which may be because students had gained experience or were
modifying programs written earlier.

Most of the syntax errors we found involved missing or mismatched parentheses.
Other common errors included writing \lstinline|else| instead of
\lstinline|else if|, using an assignment operator directly before a block
statement, and redeclared \lstinline|let|-bound variables.

\subsection{Estimation of Errors Caught by RoboJS}
\label{rt}

A post hoc analysis of the errors caught by RoboJS is less straightforward,
because these programs cannot run without a (simulated) robot and the playing
field in a particular configuration. Therefore, instead of counting the number
of errors exactly, we conservatively estimate them by searching the text of all
revisions for patterns that fail in RoboJS. For example, it is straightforward
to check for arity-mismatch errors that involve RoboJS library functions,
because we know their arity beforehand. In all, we wrote 340 regular
expressions that represent patterns of code that RoboJS rejects and manually
audited all matches to verify that they were valid. This approach is certain to
produce false negatives and thus undercounts the effectiveness of RoboJS. On
the other hand, false positives would only occur if students had unreachable
code. In summary, 19\% of the revisions had errors that RoboJS catches
and JavaScript does not.

\Cref{js-v-robojs} breaks down the errors that RoboJS catches by category.
The table has an example for each category, describes how JavaScript ordinarily
behaves and the consequences of its behavior. In almost all cases, RoboJS
catches the error early and halts the program with a sensible error message,
instead of silently failing, which is what ordinary JavaScript does.

Note that arity mismatch errors account for two-thirds of all errors caught by
RoboJS. Moreover, two library methods accounted for nearly all the arity
mismatch errors (particularly for account 10). As such, this suggests these
methods were not properly explained or have misleading names, and thus
presents an opportunity for improvement in the future.

To summarize, we find that when students ran their programs, they failed
with syntax errors 10\% of the time and with RoboJS errors 19\% of the time.
These RoboJS failures are helpful to students, because without RoboJS,
JavaScript simply fails silently and behaves in unintuitive ways. Therefore,
we conclude that RoboJS helped students catch errors earlier than if we had
used ordinary JavaScript.

\begin{table}
  \centering
  \scriptsize
  \begin{tabular}{|p{1.8cm}|r|p{4.6cm}|}\hline
    \thead{Pitfall} & \thead{Count} & \thead{Contrasting JavaScript and
    RoboJS}\\\hline
    Loose comparison\newline \textbf{Ex:} \texttt{true == 1} & 18 &
    \textbf{JS Behavior:} Operand type coercion.\newline \textbf{Consequence
    in JS:} Potential false positives.\newline \textbf{RoboJS Behavior:} Not
    allowed.\\\hline
    Uninitialized variable\newline \textbf{Ex:} \texttt{let x;} & 21 &
    \textbf{JS:} \texttt{undefined} reference.\newline \textbf{Consequence:}
    Propagation of \texttt{undefined} (if referenced before initialization).
    \newline \textbf{RoboJS:} Not allowed.\\\hline
    Conditional assignment\newline \textbf{Ex:} \texttt{if(x = 0)} & 28 &
    \textbf{JS:} Branches on value of RHS as a Boolean.\newline
    \textbf{Consequence:} Potential branching based on non-Boolean literals.
    \newline \textbf{RoboJS:} Same branch behavior, but only allowed if RHS
    evaluates to \texttt{true} or \texttt{false}.\\\hline
    Op type mismatch\newline \textbf{Ex:} \texttt{`x' * 2} & 42 &
    \textbf{JS:} \texttt{NaN} reference in the case of arithmetic/bitwise
    operators and shorthand assignment.\newline
    \textbf{Consequence:} Potential propagation of \texttt{NaN}.\newline
    \textbf{RoboJS:} Not allowed.\\\hline
    Arity mismatch\newline \textbf{Ex:} \texttt{setX();} & 213 &
    \textbf{JS:} Treats elided parameters as \texttt{undefined}; silently
    discards extra parameters.\newline \textbf{Consequence:} Propagation of
    \texttt{undefined} depending on function implementation.\newline
    \textbf{RoboJS:} Only allowed for \emph{native} JS functions.\\\hline
  \end{tabular}
  \caption{A comparison of behaviors with respect to a handful of beginner
  programming pitfalls.}
  \label{js-v-robojs}
\end{table}

\section{Future Work and Conclusion}
\label{conclusion}

With some technological improvements and feature development, we plan to
continue doing similar outreach workshops to further evaluate our methodology
effectiveness.

It is also interesting to explore other avenues to utilize our existing
platform. One idea is to use our system for future RoboCup team tryouts.
Potential aspiring team members could have the opportunity to showcase their
programming abilities and understanding of robotics principles through this
higher-level medium. Should they exhibit proficiency in our environment, they
can move onto the actual low-level C++ development necessary for RoboCup. A
second possibility would be the use of RoboJS in other teaching scenarios.
Taking away the robotics library from RoboJS, reveals a strict subset of
JavaScript that removes the hard edges of the language. This could be used for
instructional purposes, and even general use, for better JavaScript programs
and developer practice.

In conclusion, we present a means for teaching introductory computing and
robotics principles with an industry-standard programming language and
high-performance robots. This is made possible by careful learning
abstractions coupled with safe and robust technology interfaces that are made
available in a conducive environment.

\section{Acknowledgments}
\label{Acknowledgments}

This work was partially supported by NSF Awards CCF-1717636 and IIS-1724101.
Additional support was received from AFRL and DARPA agreement FA8750-16-2-0042.
We thank Jarrett Holtz, Emily Pruc, and Sadegh Rabiee for their support in
workshop preparation. We thank Emily Herbert and Donald Pinckney for their help
running the workshop. We thank Andrew Pasquale and the HolyokeCodes team for
their extensive help.

\fontsize{9.0pt}{10.0pt} \selectfont
\bibliographystyle{aaai}
\bibliography{main}

\end{document}